\documentclass[prb,twocolumn,showpacs]{revtex4}
\usepackage{graphicx}

\newcommand{\UP}{n_{\uparrow}}
\newcommand{\DN}{n_{\downarrow}}
\newcommand{\TUP}{\tau_{\uparrow}}
\newcommand{\TDN}{\tau_{\downarrow}}

\newcommand{\be}{\begin{equation}}
\newcommand{\ee}{\end{equation}}
\newcommand{\bea}{\begin{eqnarray}}
\newcommand{\eea}{\end{eqnarray}}
\newcommand{\bean}{\begin{eqnarray*}}
\newcommand{\eean}{\end{eqnarray*}}

\begin{document}

\title{Dimensional crossover of the exchange-correlation energy at the semilocal 
level}
\author{Lucian A. Constantin}
\affiliation{ Department of Physics and
Quantum Theory Group, Tulane University, New Orleans, LA 70118}

\date{\today}

\begin{abstract}
Commonly used semilocal density functional approximations for the exchange-correlation energy 
fail badly 
when the true two dimensional limit is approached. We show, using a quasi-two-dimensional uniform 
electron gas in the infinite barrier model, that the 
semilocal level can 
correctly recover the exchange-correlation energy of the two-dimensional uniform 
electron gas.
% and can be accurate in the quasi-two-dimensional uniform electron gas 
% region. 
We derive new exact constraints at the semilocal level 
for the dimensional crossover of the exchange-correlation energy and we propose a method to 
incorporate them in any exchange-correlation density functional approximation.

\end{abstract}


\maketitle

\section{Introduction}
\label{sec1}
\noindent

In the Kohn-Sham time-independent density functional theory (DFT) the noninteracting
kinetic energy is treated as an exact functional of the occupied
orbitals~\cite{KS} and only the exchange-correlation (xc) energy $E_{xc}$
has to be approximated. The "Jacob's ladder" classification \cite{jacob} of the ground-state 
density-functional
approximations for $E_{xc}$ has three complete non-empirical rungs:
the local-spin-density approximation (LSDA)~\cite{KS}, the generalized gradient
approximation (GGA)~\cite{PBE,PBEsol}, and the meta-GGA~\cite{TPSS}. Higher rungs of the ladder 
require new ingredients in order to satisfy more exact constraints \cite{jacob}. Thus, the meta-GGA 
has as 
ingredients
the spin densities $\UP$ and $\DN$, their gradients $\nabla\UP$ and
$\nabla\DN$,
and the Kohn-Sham (KS) noninteracting kinetic energy densities $\TUP$ and $\TDN$. 
The local and semilocal density
functionals (LSDA, GGA, and meta-GGA) give accurate predictions of 
ground-state for 
atoms, molecules, solids ~\cite{KPB,SSTP}, and
surfaces~\cite{Con}. They also work for atomic monolayers~\cite{RAM,VRS}
and other quasi-two-dimensional (quasi-2D) systems~\cite{RM}, but they fail badly as the true 2D limit 
is
approached~\cite{PoP,KLNLHM}. The failure of the semilocal density functionals to describe
the dimensional crossover of the exact xc functional can be avoided by using
nonlocal models \cite{GGo,GGG} 
such as the weighted density approximation \cite{GJL},
or higher rungs of the "Jacob's ladder" \cite{CPP}. 
Thus, the fourth-rung hyper-GGA, a nonlocal correlation functional compatible with exact 
exchange \cite{Perdew},  improves considerably the behaviour of semilocal functionals over the whole 
thickness 
range of the quasi-2D electron gas \cite{CPP}. The numerically expensive fifth-rung approximations such 
as the inhomogeneous 
Singwi-Tosi-Land-Sj\"olander method (ISTLS) \cite{DWG} and the GW approximation \cite{Hedin} 
are remarkably accurate for the description of quasi-2D systems \cite{CPP,GGG}.

The quasi-2D electron gas is experimentally realizable in silicon metal-oxide-semiconductor 
field-effect 
transistor (MOSFET) and in the widely used semiconductor heterojunctions \cite{KLNLHM,BPE}. 
Other physical systems with strong 2D character are the copper-oxide planes of
high-temperature superconductors and the electrons bound to the surface of liquid helium \cite{BPE}.
%In this work we show that the dimensional crossover of the xc energy can be accurately 
%described 
%at the meta-GGA level, thus the computational attractive semilocal functionals can be used 
%for calculations of physical systems with strong 2D character.

The paper is organized as follows. In section \ref{sec2},
we present the exact constraints at the semilocal level for the dimensional crossover of 
the exchange-correlation energy. In section \ref{sec3}, we construct a simple 
semilocal functional that incorporates these exact conditions, and we test it for the 
quasi-2D uniform electron gas, jellium slabs and non-uniformly-scaled hydrogen atom.
In section \ref{sec4}, we summarize our
conclusions.

%\section{EXACT CONSTRAINTS FOR THE DIMENSIONAL CROSSOVER OF THE EXCHANGE-CORRELATION ENERGY}
\section{EXACT CONDITIONS FOR SEMILOCAL DENSITY FUNCTIONALS}
\label{sec2}

\noindent

A 2D uniform electron gas is described by the 2D electron-density parameter
$r_s^{2D}=1/\sqrt{\pi n^{2D}}=\sqrt{2}/k_F^{2D}$.
(Unless otherwise stated, atomic units are used throughout, i.e.,
$e^2=\hbar=m_e=1$.)
Here $n^{2D}$ is the density
of electrons per unit area, and $k_F^{2D}$ represents the magnitude of the
corresponding 2D Fermi wavevector. The exchange energy per particle of the 2D uniform electron gas is 
\cite{SPL}
\begin{equation}
\epsilon^{2D}_x=-(4\sqrt{2}/(3\pi))/r_s^{2D}=-0.6002/r_s^{2D}.
\label{e1}
\end{equation}
The correlation energy per particle of a 2D uniform electron gas in the high-density limit 
($r_s^{2D}\rightarrow 0$) is \cite{SPL}
\begin{equation}
\epsilon^{2D}_c=-0.19-0.086r_s^{2D}\ln
r_s^{2D}+O(r_s^{2D}),
\label{e2}
\end{equation}
and in the low-density limit
($r_s^{2D}\rightarrow\infty$) is \cite{SPL}
\begin{equation}
\epsilon^{2D}_c=(\frac{8}{3\pi}-2+\frac{4\sqrt{2}}{3\pi})(r_s^{2D})^{-1}+(r_s^{2D})^{-3/2}+
O((r_s^{2D})^{-2}).
\label{e3}
\end{equation}
A realistic interpolation (which uses Quantum Monte Carlo data) between the high- and
low-density limits of the 2D correlation energy per electron has the following form~\cite{PoP,SPL}
\begin{equation}
\epsilon^{2D}_c=0.5058\left[\frac{1.3311}{(r^{2D}_s)^2}\left(\sqrt{1+1.5026
r^{2D}_s}-1\right)-\frac{1}{r^{2D}_s}\right].
\label{e4}
\end{equation}

Similarly to Ref. \cite{CPP}, 
let us consider a quantum well of thickness $L$ in the $z$-direction.
In the infinite-barrier model (IBM) \cite{Ne} for a quantum well, the KS effective one-electron 
potential is
zero inside the well and infinity outside it, such that the KS orbital is
\begin{equation}
\Psi_{l,k}=\sqrt{\frac{2}{AL}}\sin(\frac{l\pi z}{L})e^{i\bf{r}_{||}\bf{k}_{||}}
\;\;\;\;\rm{for}
\;\;\;\;0\leq z\leq L,\;\;l\geq 1,
\label{e5}
\end{equation}
where $A$ is the area of the $xy$-plane, $l$ is the subband index, and
$\bf{r}_{||}$ and
$\bf{k}_{||}$ are the position and the wavevector parallel to the surface.
In this model the electrons can not leak out of the well,
so the true 2D electron-gas limit is
recovered by shrinking the well. The energy
levels of this model are \cite{PoP}:
\begin{equation}
E_{l,k}=\frac{1}{2}[(\frac{l\pi}{L})^2+k_{||}^2].
\label{e6}
\end{equation}
When only the lowest level is occupied ($E_{1,k_F^{2D}} < E_{2,0}$ which
implies
$L<\sqrt{3/2}\pi r_s^{2D}=L_{\rm{max}}$ \cite{PoP}), the density of states of
this system begins to resemble the density of states of a 2D electron gas, the
motion in the
z-direction is frozen out, and the system can be considered quasi-two-dimensional.

By shrinking the $z$-coordinate without changing the total number of electrons per unit area, the
system reaches the 2D electron gas limit. This process is equivalent with
a non-uniform scaling in one dimension \cite{LO}, and the 3D scaled
density is \cite{PoP}
\begin{equation}
n^z_{\lambda}(z)=\frac{2}{(L/\lambda)\pi(r_s^{2D})^2}\sin^2(\frac{\pi
z}{(L/\lambda)})\; ; \;\;\;\;0\leq
z \leq L/\lambda,
\label{e7}
\end{equation}
where $n^z_{\lambda}(x,y,z)=\lambda n(x,y,\lambda z)$ and $\lambda$ is the scaling parameter.
When $\lambda\rightarrow\infty$, $L/\lambda<<L_{\rm{max}}$, and the 2D limit is achieved.
The corresponding exchange and correlation energies per particle should satisfy the 
following scaling
relations \cite{PoP}
\begin{equation}
\lim_{\lambda\rightarrow\infty}\frac{1}{N}E_x[n^x_{\lambda}]>-\infty\;\;;\;\;\;
\lim_{\lambda\rightarrow\infty}\frac{1}{N}E_c[n^x_{\lambda}]>-\infty,
\label{e8}
\end{equation}
where $N=\int^L_0 n(z)dz$. These equations, which start from those of Ref.~\cite{LO}, are not
satisfied by LSDA, GGA or meta-GGA.

The GGA exchange-correlation energy per particle of our quantum well of thickness $L$
is
\begin{equation}
\frac{E^{GGA}_{xc}}{N}=(\int^{L}_{0}n(z)\epsilon^{GGA}_{xc}(n(z),\nabla
n(z))dz)/N,
\label{e9}
\end{equation}
and the meta-GGA  exchange-correlation energy per particle is
\begin{eqnarray}
&&\frac{E^{MGGA}_{xc}}{N}=(\int^{L}_{0}n(z)\epsilon^{MGGA}_{xc}(n(z),\nabla
n(z),\tau(z))dz)\cr\cr
&/& N,
\label{e10}
\end{eqnarray}
where $\epsilon^{GGA}_{xc}$ and $\epsilon^{MGGA}_{xc}$ are the GGA and meta-GGA  xc
energies per particle of the 3D system.

The positive kinetic energy density of the IBM quasi-2D electron gas is 
\begin{equation}
\tau=\tau^W+\frac{(k_F^{2D})^4}{4\pi (L/\lambda)}\sin^2(\frac{\pi z}{(L/\lambda)})\geq\tau^W,
\label{e11}
\end{equation}
where $\tau^W=\pi (k_F^{2D})^2/(2 (L/\lambda)^3) \cos^2(\frac{\pi z}{(L/\lambda)})$ is the von
Weizs\"{a}cker
kinetic energy density \cite{vW}. 
When $\lambda\rightarrow\infty$, $\tau\rightarrow\tau^W\sim\lambda^3$ and $\tau-\tau^W\sim\lambda$.
Eq. (\ref{e11}) can be well described by the Laplacian-level 
meta-GGA kinetic energy density of Ref. \cite{PC3}.

The reduced gradients for exchange ( $p=|\nabla n|^2/[4(3\pi^2)^{2/3}n^{8/3}]\sim 
\lambda^{4/3}$) measures the variation of the density over a Fermi wavelength and that for correlation 
( 
$t=|\nabla n|/[4(3/\pi)^{1/6}n^{7/6}]\sim\lambda^{5/6}$) measures the variation of the density over 
the screening length. Both
tend to infinity when the 2D limit is reached ($\lambda\rightarrow\infty$), such that in the 
quasi-2D electron gas regime, the density is rapidly varying almost everywhere. 
Thus this 
system is not only a challenge for a semilocal density functional, but it can give 
exact constraints (at the semilocal level) in the regime where the 3D density and 
its gradient diverge.  

The exact exchange energy per particle of the 2D uniform electron gas (see Eq. (\ref{e1}))
can be \emph{correctly} recovered by any 3D semilocal density 
functional approximation for exchange energy
if in the large gradient limit ($p\rightarrow\infty$)
the 3D exchange energy per particle behaves as $a_x p^{-1/4}\epsilon_x^{LSDA}$ (see 
Ref.\cite{note3}),
where 
\begin{equation}
a_x=(0.6002 \sqrt{2}\pi)/(3^{3/2}\int^1_0 dy \sin^{7/2}(\pi y)\cos^{-1/2}(\pi y)),
\label{e12}
\end{equation}
was derived in the IBM model using Eqs. (\ref{e1}) and (\ref{e9}). Eq. (\ref{e12}) gives
$a_x=0.5217$. The parameter $a_x$ (as well as the other results of this section), even if 
calculated using the IBM quasi-2D electron gas, is 
independent on the quantum well potential model along the confinement $z$-direction because the 2D 
limit ($\lambda\rightarrow\infty$) is not relaying on the effective potential model.

At a meta-GGA level, we can also use the dimensionless inhomogeneity parameter \cite{PTSSex,TPSS} 
\begin{equation}
\alpha=\frac{\tau-\tau^W}{\tau^{unif}}\sim\lambda^{-2/3},
\label{e13}
\end{equation}
where $\tau^{unif}=(3/10)(3\pi^2)^{2/3}n^{5/3}$ is the Thomas-Fermi kinetic energy 
density of the 3D uniform electron gas \cite{TF}.
Thus Eq. (\ref{e1}) can also be \emph{exactly}
satisfied by any meta-GGA if in the large gradient limit ($p\rightarrow\infty$) 
the 3D exchange energy per particle behaves as $b_x \alpha^{1/2}\epsilon_x^{LSDA}$,
where $b_x=1.947$ was found similarly as $a_x$, from Eqs. (\ref{e1}) and (\ref{e10}).

Because $r_s^{2D}$ enters in a nonlinear manner in Eq. (\ref{e4}), the GGA level can not 
describe
the correlation energy per particle of a 2D uniform electron gas, but it can explain the 2D 
high- and
low-density limits. The high-density limit of Eq. (\ref{e2}) can be exactly recovered by 
any 3D GGA that behaves in the large gradient limit ($t\rightarrow\infty$) as
\begin{eqnarray}
\epsilon^{GGA}_c\rightarrow 
-0.19-0.0497n^{-5/12}
\sqrt{t}\ln(n^{-5/12}\sqrt{t}),
\label{e14}
\end{eqnarray}
where $t\sim\lambda^{5/6}$ is the reduced gradient for correlation.
The low-density limit of Eq. (\ref{e3}) can also be exactly recovered by 
any 3D GGA that behaves in the large gradient limit ($t\rightarrow\infty$) as
\begin{eqnarray}
\epsilon^{GGA}_c\rightarrow
-0.40345n^{5/12}t^{-1/2} +0.459 n^{5/8}t^{-3/4}.
\label{e15}
\end{eqnarray}

At the meta-GGA level, Eq.(\ref{e4}) can be correctly satisfied
by any 3D meta-GGA that behaves in the large gradient limit ($t\rightarrow\infty$) as 
\begin{eqnarray}
\epsilon^{MGGA}_c\rightarrow \epsilon_c^{2D},
\label{e16}
\end{eqnarray}
where $\epsilon_c^{2D}$ is given by Eq.(\ref{e4}) and
\begin{equation}
r_s^{2D}=0.4173\; n^{-1/3}\;\alpha^{-1/2}=0.6727\; r_s^{3D}\;\alpha^{-1/2}.
\label{e17}
\end{equation}
Eq. (\ref{e17}) connects $r_s^{2D}$ with $r_s^{3D}$, showing the importance of the $\alpha$ 
ingredient
to the dimensional crossover of the exchange-correlation energy, and to the non-uniform scaling 
in one dimension.

%\section{EXACT CONSTRAINTS FOR THE DIMENSIONAL CROSSOVER OF THE EXCHANGE-CORRELATION ENERGY}
\section{CONSTRUCTION AND TESTS OF A SIMPLE SEMILOCAL FUNCTIONAL}
\label{sec3}

\noindent

The results of Section \ref{sec2} can be included in any density functional approximation for 
the
exchange-correlation energy. For simplicity, we incorporate them in the LSDA. Let consider 
first the exchange part, and define GGA+2D and meta-GGA+2D as
\begin{equation}
\epsilon^{GGA+2D}_x=\epsilon^{LSDA}_x\{1+f(p)[-1+0.5217p^{-1/4}]\},
\label{e18}
\end{equation}
and
\begin{equation}
\epsilon^{MGGA+2D}_x=\epsilon^{LSDA}_x\{1+f(p)[-1+1.947\alpha^{1/2}]\},
\label{e19}
\end{equation}
where $f(p)=1$ for $p=\infty$.
The simplest approximation for $f(p)$ is a step function
\begin{equation}
f(p)=\lim_{x->\infty} \theta(p-x),
\label{e20}
\end{equation}
where $\theta(x)$ is $0$ for $x< 0$ and $1$ for $x\geq 0$.
This model preserves all the exact constraints that the local or semilocal functional
satisfies, and recovers the exchange energy of the 2D uniform electron gas in the 
limit $L=0$ (when $p=\infty$). However, this approximation does not improve the 
behavior of the semilocal functional in the quasi-2D region, and moreover, it gives a 
discontinuity when $p\rightarrow\infty$.

We propose the following simple analytic model for the function $f(p)$ \cite{note1}:
\begin{equation}
f(p)=\frac{p^4(1+p^2)}{10^c+p^6}.
\label{e21}
\end{equation}
where $c>0$ is an empirical parameter. Eq. (\ref{e21}) recovers the right limit when 
$p\rightarrow\infty$ ($f(p\rightarrow\infty)=1$), and for a slowly-varying density, when $p$ 
is small
($p<1$), it behaves as $f(p)\sim [p^4 10^{-c}+\rm{higher\; order \;terms}]$. This 
is a good feature because it can accurately preserve the behavior of the semilocal 
functional in the slowly-varying limit \cite{note2}. 
When $c$ is large, Eq. (\ref{e21}) 
starts to model Eq. (\ref{e20}).    

\begin{figure}
\includegraphics[width=\columnwidth]{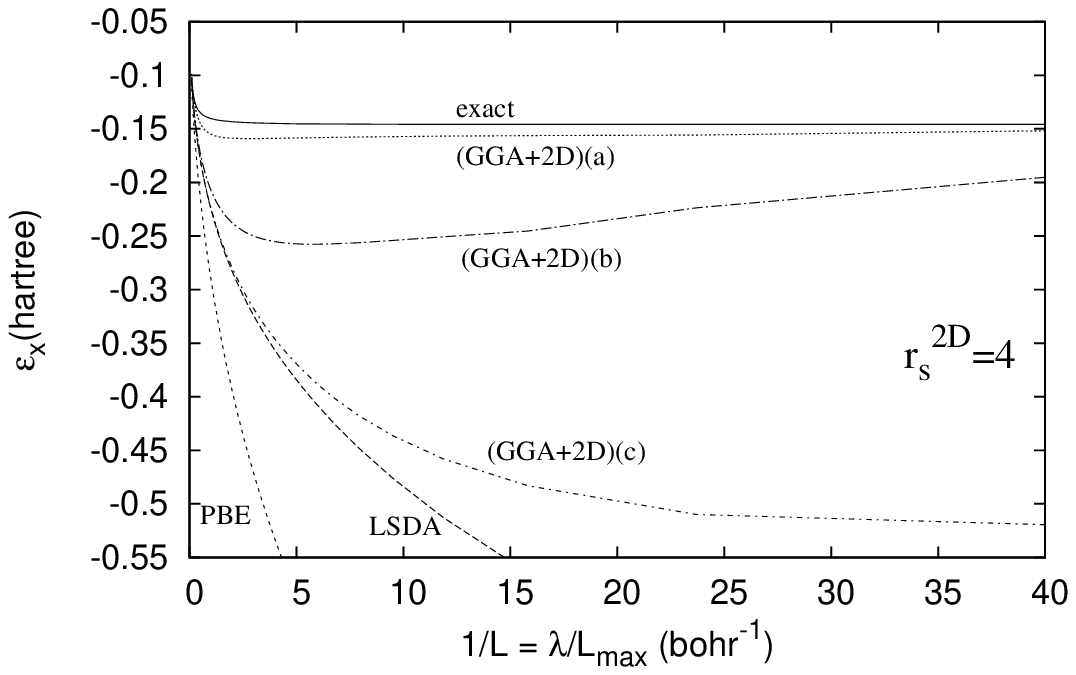}
\includegraphics[width=\columnwidth]{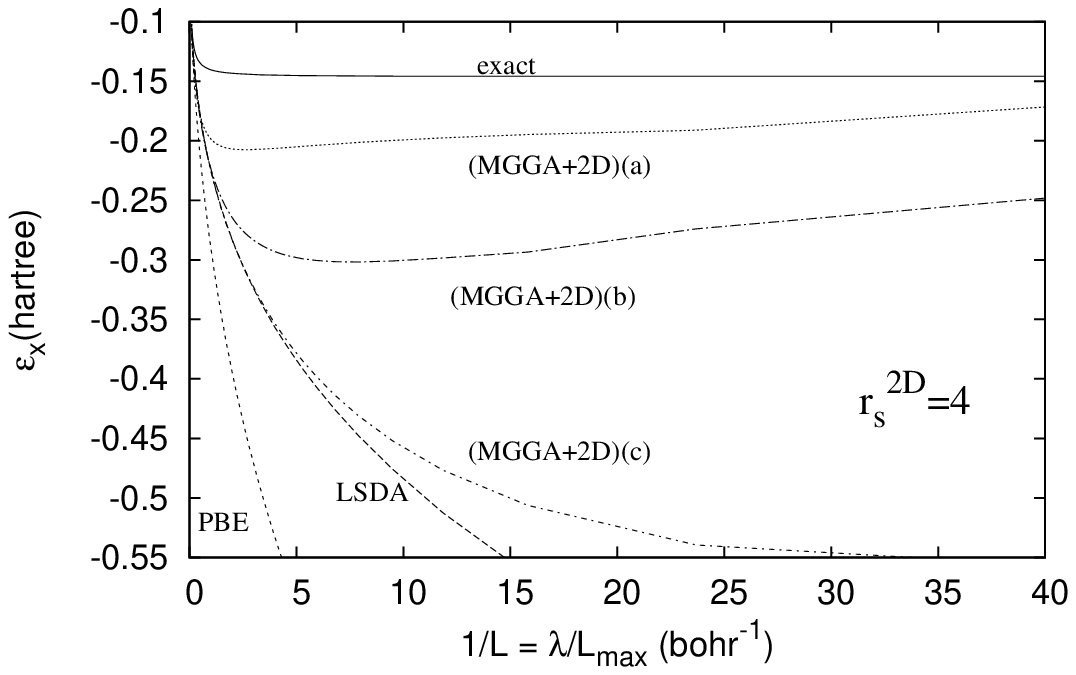}
\caption{ Exchange energy per particle of an IBM quasi-2D electron
gas of fixed 2D electron density ($r_s^{2D}=4$), as a function of the
inverse quantum well thickness $1/L$ ($L<L_{\rm{max}}=15.39$).
The curves denoted by (GGA+2D)(a), (GGA+2D)(b), and (GGA+2D)(c) are those 
given by Eq. (\ref{e18}), using for $f(p)$ the analytic model of 
Eq. (\ref{e21}), with $c=2$, $8$ and $16$ respectively.
The curves denoted by (MGGA+2D)(a), (MGGA+2D)(b), and (MGGA+2D)(c) are those
given by Eq. (\ref{e19}), using for $f(p)$ the analytic model of
Eq. (\ref{e21}), with $c=2$, $8$ and $16$ respectively. 
While the LSDA and PBE diverge, all GGA+2D and MGGA+2D curves 
recover in the limit $\lambda\rightarrow\infty$
the corresponding exchange energy (-0.15005) of a 2D electron gas.
}
\label{f2}
\end{figure}

In Fig. \ref{f2}, we show the exchange energy per particle
of the quasi-2D electron gas with
2D bulk parameter $r_s^{2D}=4$ for
several density functionals:
exact exchange, LSDA, PBE GGA \cite{PBE}, GGA+2D of Eq. (\ref{e18}) and MGGA+2D of Eq. 
(\ref{e19}), using in Eq. (\ref{e21}) three values for the parameter $c$ ($c=$2, 8, and 16).
LSDA and PBE
diverge when $L\rightarrow 0$ ($\lambda\rightarrow\infty$).
The meta-GGA TPSS \cite{TPSS}, not plotted in Fig. \ref{f2}, has the same behavior as PBE.
Both GGA+2D and MGGA+2D perform
better than LSDA at small thicknesses of the quantum well, and both of them recover the
exact exchange energy of the 2D uniform electron gas when $L=0$.
However, we observe that GGA+2D with $c=2$ (the curve denoted (GGA+2D)(a)) is the most accurate, 
and remarcably describes the quasi-2D region. When the value of parameter $c$ increases, 
the GGA+2D and meta-GGA+2D have the LSDA behavior over a larger region, and consequently,
they are not accurate in the quasi-2D region. We also remark that  GGA+2D and meta-GGA+2D 
calculated with $c=8$ (the curves (GGA+2D)(b) and (MGGA+2D)(b)) give a significant improvement 
over the LSDA in the whole quasi-2D region.
%
%%%%%%%%%%%%%%%%%%%%%%%%%%%%%%%%%%%%%%%%%%%%%%%%%%%%%
%%%%%%%%%%%%%%%%%%%%%%%%%%%%%%%%%%%%%%%%%%%%%%%%%%%%%%%
%

\begin{figure}
\includegraphics[width=\columnwidth]{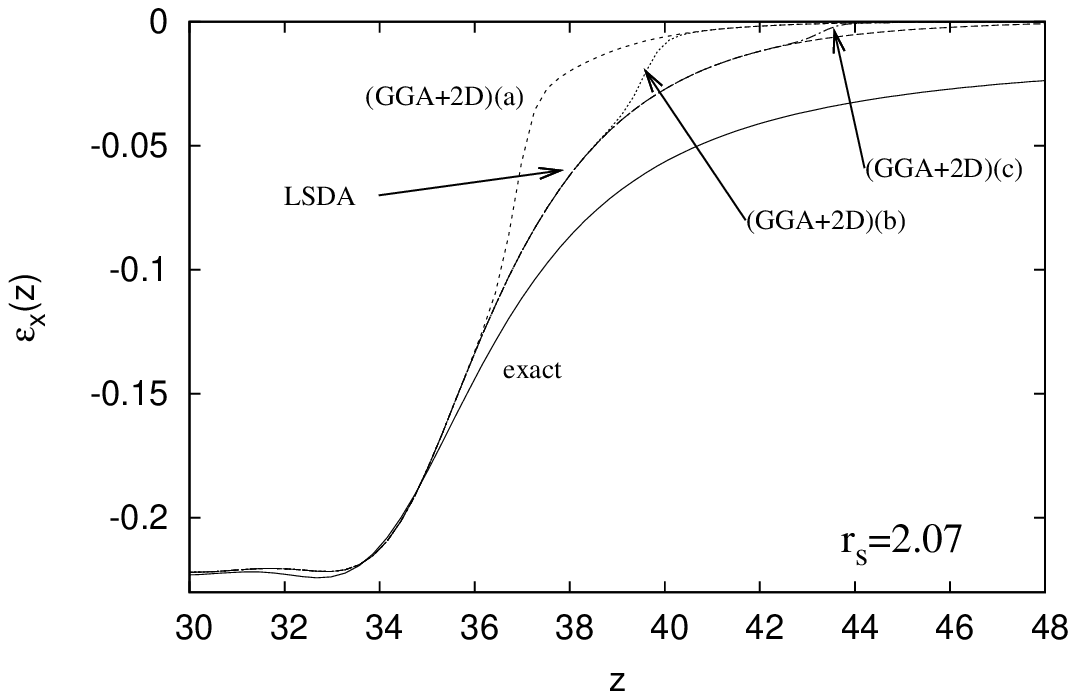}
\includegraphics[width=\columnwidth]{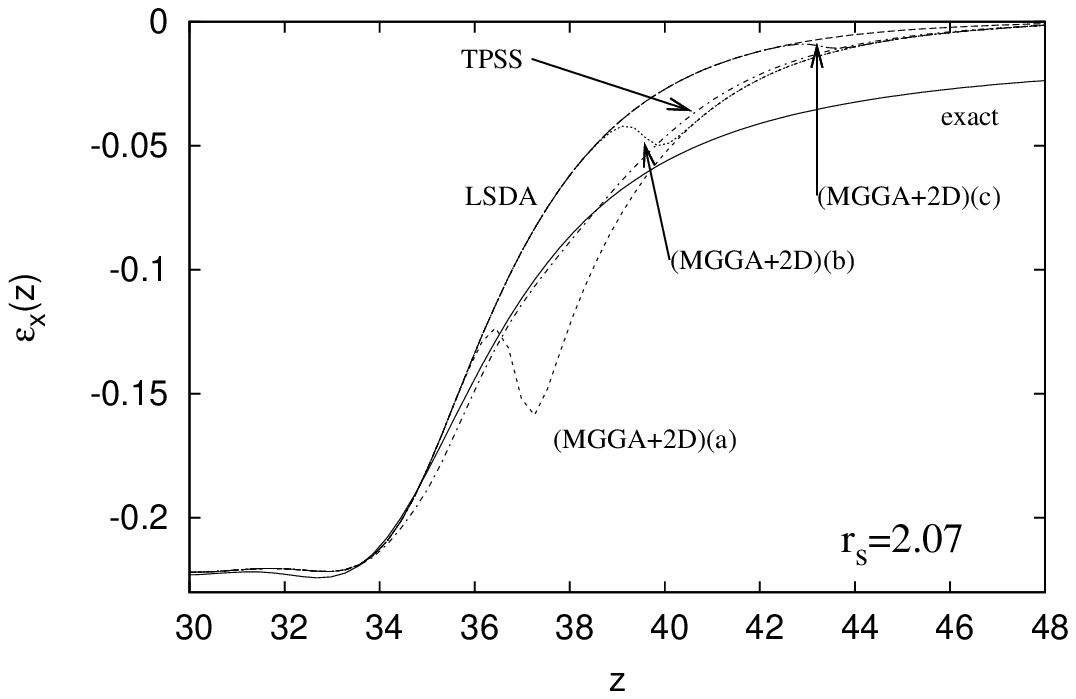}
\caption{ Exchange energy per particle at position $z$
versus $z$ in atomic units, at a jellium slab surface.
The bulk parameter is $r^{3D}_s=2.07$, the slab width is $d=3.2\lambda_F$
 and the jellium surface
is at $z=35.24 a.u.$ ($\lambda_F=2\pi/k^{3D}_F$ is the Fermi wavelength). The calculations of the exact 
exchange, LSDA, 
TPSS, GGA+2D and 
MGGA+2D
use the LSDA Kohn-Sham orbitals. The curves (GGA+2D)(a), (GGA+2D)(b), (GGA+2D)(c), and
(MGGA+2D)(a), (MGGA+2D)(b), (MGGA+2D)(c) has the same signification as in Fig. \ref{f2}.}
\label{f3}
\end{figure}

In Fig. \ref{f3}, we show the exchange energy per particle of a thick jellium slab of
bulk parameter $r^{3D}_s=2.07$. (The bulk parameter defined by the equation $n=3/4\pi (r^{3D}_s)^3$, 
represents the radius of a sphere that encloses on average one electron). 
The local and semilocal density 
approximations 
(LSDA, TPSS, GGA+2D, 
and 
MGGA+2D) show an exponential
decay of the exchange energy per particle whereas the exact exchange behaves as
$\approx -1/(4z)$ \cite{SS3}. 
All the  MGGA+2D curves have a bump in the region where $f(p)$ switches from 0 to 1, and after  
that they are close to the TPSS 
meta-GGA exchange energy per particle. (We recall that TPSS meta-GGA has the same large-$p$ 
behavior as the PBE-GGA). 
We observe that for jellium slabs (as well as for many 3D systems) $E^{GGA+2D}_x>E^{LSDA}_x$, 
whereas $E^{MGGA+2D}_x<E^{LSDA}_x$, thus one can try 
also a convex combination between Eq.(\ref{e18}) and Eq.(\ref{e19}). However, the construction 
of an 
accurate 3D and quasi-2D semilocal functional is a difficult task \cite{note}, and is beyond 
the purpose of this paper.  
%%%%%%%%%%%%%%%%%%%%%%%%%%%%%%%%%%%%%%%%%%%%%%%%%%%%%
%%%%%%%%%%%%%%%%%%%%%%%%%%%%%%%%%%%%%%%%%%%%%%%%%%%%%%%

Since $p$ values bigger than 9 are found in
the tail of an atom or molecule, where the electron density is negligible, we can 
choose $c=8$ (such that at $p=9$, $f=0.0053$ and for $p>50$, $f\rightarrow 1$).
This choice ensures that the GGA+2D of Eq. (\ref{e18}) and meta-GGA+2D of 
Eq. (\ref{e19}) perform similarly 
with LSDA for 
3D systems and make a considerable improvement in quasi-2D region, and recover the 
exchange energy of the 2D uniform electron gas. 
Thus, from now, all the presented calculations use $c=8$ in Eq. (\ref{e21}). 

Similar with the exchange case, the correlation results of Section \ref{sec2} 
can be included in any density functional approximation, however, for simplicity
we again show them for the LSDA case. Thus, we define
\begin{equation}
\epsilon^{MGGA+2D}_c=\epsilon^{LSDA}_c+f(t)[-\epsilon^{LSDA}_c+\epsilon^{2D}_c],
\label{e22}
\end{equation}
where $t$ is the reduced gradient for correlation, $f(t)$ has the same form as Eq. (\ref{e21}) 
(with $c=8$), and $\epsilon^{2D}_c$ is given by Eqs. (\ref{e4}) and (\ref{e17}).

\begin{figure}
\includegraphics[width=\columnwidth]{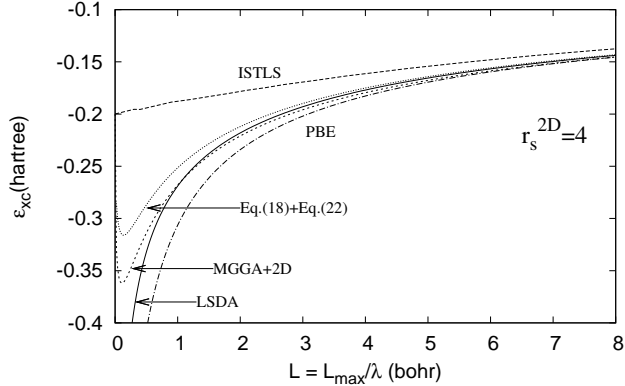}
\caption{ Exchange-correlation energy per particle of an IBM quasi-2D electron
gas of fixed 2D electron density ($r_s^{2D}=4$), as a function of the
quantum well thickness $L$ ($L<L_{\rm{max}}=15.39$).
While LSDA and PBE diverge in the 2D limit, MGGA+2D and $\epsilon^{GGA+2D}_x+\epsilon^{GGA+2D}_c$ (see 
Eqs. (\ref{e18}) and (\ref{e22})) approach the exact 2D limit. 
}
\label{f5}
\end{figure}

\begin{figure}
\includegraphics[width=\columnwidth]{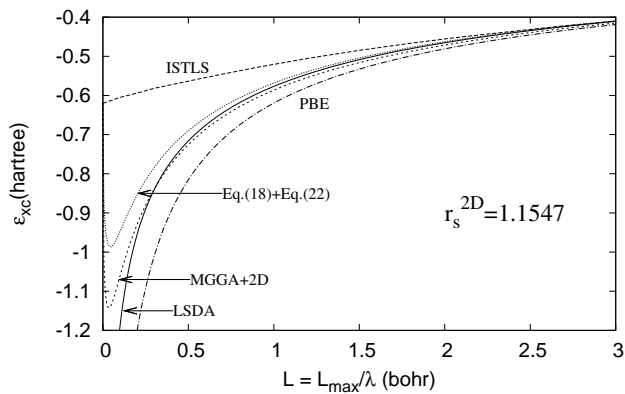}
\caption{ 
As in Fig. \ref{f5}, but now for $r_s^{2D}=2/\sqrt{3}$ ($L_{\max}=4.44$).
}
\label{f6}
\end{figure}

Figures \ref{f5} and \ref{f6} show 
several approximations of the xc energy per particle versus the quantum-well thickness $L$,
%
%the xc energy per particle of several approximations
for quasi-2D electron gases of fixed 
2D electron-density parameters: $r_s^{2D}=4$ and $r_s^{2D}=2/\sqrt{3}$ (as Figs. 1 and 2 of Ref. 
\cite{CPP}). The ISTLS method \cite{DWG}, a self-consistent approach that depends on all occupied and 
unoccupied KS orbitals,  
is remarkably accurate for any thickness $L=L_{\max}/\lambda$ (see 
Ref.\cite{CPP}). LSDA and PBE are accurate in the limit $L\rightarrow L_{\max}$, but they fail badly 
in the zero thickness limit. MGGA+2D and the xc energy per particle of Eqs. (\ref{e18}) and 
(\ref{e22}) ($\epsilon^{GGA+2D}_x+\epsilon^{MGGA+2D}_c$) 
are accurate in the limit $L\rightarrow L_{\max}$ and improve considerably the behavior of LSDA when 
$L<\sim 0.5 L_{\max}$, approaching the exact 2D limit when $L\rightarrow 0$.       

%%%%%%%%%%%%%%%%%%%%%%%%%%%%%%%%%%%%%%%%%%%%%%%%%%%%%
%%%%%%%%%%%%%%%%%%%%%%%%%%%%%%%%%%%%%%%%%%%%%%%%%%%%%%%

%%%%%%%%%%%%%%%%%%%%%%%%%%%%%%%%%%%%%%%%%%%%%%%%%%%%%
%%%%%%%%%%%%%%%%%%%%%%%%%%%%%%%%%%%%%%%%%%%%%%%%%%%%%%%

\begin{figure}
\includegraphics[width=\columnwidth]{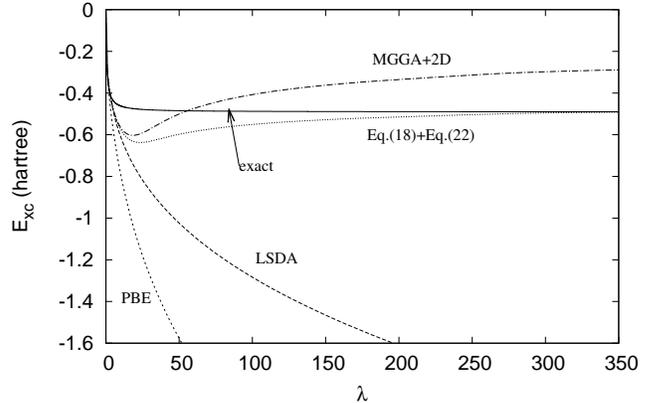}
\caption{
Exchange-correlation energy versus scaling parameter $\lambda$
for the non-uniformly-scaled hydrogen atom. Also shown is the 
exact xc energy (see Eq. (\ref{e23})).
}
\label{f7}
\end{figure}

Let us present our results for the non-uniformly-scaled hydrogen
atom \cite{PC3,Kurth}, whose density is $n_{\lambda}(\mathrm{r})=(\lambda/\pi)\exp(-2\sqrt
{x^{2}+y^{2}+(\lambda z)^{2}})$.
The exact xc energy is \cite{Kurth}
\begin{equation}
E_{xc}(\lambda)=\left\{ \begin{array}{lll}
-\frac{5}{16}\frac{\lambda}{\sqrt{\lambda^2-1}}\arctan (\sqrt{\lambda^2-1}),
     & \lambda> 1\\
-\frac{5}{16},   & \lambda=1\\
-\frac{5}{16}\frac{\lambda}{\sqrt{1-\lambda^2}}\ln(\frac{\sqrt{1-\lambda^2}+1}{\lambda}), 
    & \lambda<1 .\\
                                    \end{array}
\right.
\label{e23}
\end{equation}
When $\lambda\rightarrow\infty$, this system can model an electron firmly bound to a surface.
Fig. \ref{f7} shows that LSDA and PBE fail badly in the extreme oblate case ( 
$\lambda>> 1$). 
The meta-GGA TPSS \cite{TPSS}, not plotted in Fig. \ref{f7}, has the same behavior as PBE.
Because in any one-electron system $\tau=\tau^W$ and $\alpha=0$,
the MGGA+2D xc energy will approach slowly zero in the limit $\lambda\rightarrow\infty$.
The xc energy of Eqs. (\ref{e18}) and (\ref{e22}) ($E^{GGA+2D}_x+E^{MGGA+2D}_c$) is very 
accurate at large values of $\lambda$. 
%    
%%%%%%%%%%%%%%%%%%%%%%%%%%%%%%%%%%%%%%%%%%%%%%%%%%%%%
%%%%%%%%%%%%%%%%%%%%%%%%%%%%%%%%%%%%%%%%%%%%%%%%%%%%%%%
%

In this section we have proposed 
a method that incorporates the 2D limit of the electron gas in semilocal functionals and 
keeps as much as possible the 3D accuracy of the semilocal functionals.
However, in order to obtain a good description of the quasi-2D
region, the empirical parameter $c$ has to be smaller ($c\approx 2$) than the optimized 
value ($c=8$), but such a choice
will significantly modify the 3D accuracy of the semilocal functional. In the 
investigation of 
physical systems with strong 2D character, Eq. (\ref{e17}) can be seen as 
an indicator of the quasi-2D electron gas regime. 
Various indicators of the electron localization have been constructed for 3D systems 
(see for example Ref.\cite{TVT}), but
Eq. (\ref{e17}) is a better and natural choice in the case of quasi-2D uniform gas.
Thus, in the quasi-2D regions, where 
$r^{3D}_s(z)\alpha^{-1/2}(z)$ is constant, we can choose $c=2$ and in the other regions we 
can choose $c=8$ (or even $c=\infty$). Such a parametrization of $c$ can be more useful in 
applications than the use of our optimized value for $c$ ($c=8$).

\section{CONCLUSIONS}
\label{sec4}

\noindent

In summary, we have shown that the dimensional crossover (from 3D to 2D) of the exact xc energy can be 
significantly improved at a meta-GGA level, and we derive new exact constraints (see Section 
\ref{sec2}) using an IBM quasi-2D electron gas. 
Same results can be obtained using the parabolic quantum well of Ref. \cite{KLNLHM} because the 
2D limit is independent on the quasi-2D electron gas model. 
Thus, a 3D meta-GGA that requires input from the 3D uniform electron gas \cite{PW}, can describe a 2D 
system using only the 
highly nonlocal region 
where $|\nabla 
n|\rightarrow\infty$. Moreover,
Eq. (\ref{e17}) shows a close connection between $r_s^{2D}$, $r_s^{3D}$ and the positive 
noninteracting kinetic 
energy density $\tau$ in the case of the quasi-2D electron gas.

We propose a simple approach to incorporate the dimensional crossover constraints in any local or 
semilocal density functional approximation for the xc energy, and we present it in the case 
of LSDA. 
However, future work is needed to construct an accurate meta-GGA that satisfies the dimensional 
crossover constraints.

The non-uniform scaling in one dimension is closely related to the quasi-2D electron gas. The 
non-uniformly scaled hydrogen atom in the oblate case ($\lambda\geq 1$), an important and hard test 
for the density 
functionals as well as 
a model for an electron bound to a surface, can be 
well described by our simple modified LSDAs. 
Thus we hope that this work can be useful not only for 
investigation of physical systems with strong 2D character, but also 
in developing more accurate density functionals. 

L.A.C. thanks Professor John P. Perdew and Professor J.M. Pitarke for many valuable discussions and 
suggestions.
L.A.C. acknowledges NSF support (Grant No. DMR05-01588).


\begin{thebibliography}{[Vo]}
\bibitem{note3}
This choice, found from the $L$-dependence of $n$ and $p$, is necessary and
sufficient that the exchange energy per particle of the quantum well (see Eq. (\ref{e9}))
to be finite (and nonzero) when $L=0$.
\bibitem{note1}
A smooth function that also interpolates between 0 and 1 was used 
in construction of the Laplacian-level meta-GGA kinetic energy density
(see Fig. 1 of Ref. \cite{PC3}).
\bibitem{note}
The satisfaction of Lieb-Oxford bound \cite{LO1}, that is an important constraint 
in construction of accurate density functionals, is a challenge for a GGA+2D whose 
enhancement factor tends to zero when $p$ becomes large.
\end{thebibliography}
\end{document}